\documentclass[aps,prl,twocolumn,superscriptaddress,floatfix,longbibliography]{revtex4-2}
\usepackage{amsmath,amssymb,bm,graphicx,mathrsfs}
\usepackage[hidelinks]{hyperref}
\graphicspath{{figures/}}
\AtBeginDocument{\lefthyphenmin=2 \righthyphenmin=3 }

\begin{document}

\title{Diffusion among degenerate orbits and phase timekeeping in a boundary time crystal}
\hypersetup{pdftitle={Diffusion among degenerate orbits and phase timekeeping in a boundary time crystal}}

\author{Shu He}
\email{heshu1987@foxmail.com}
\affiliation{Department of Physics and Electronic Engineering, Sichuan Normal University, Chengdu 610066, China}
\author{Liwei Duan}
\email{duanlw@gmail.com}
\affiliation{Department of Physics, Zhejiang Normal University, Jinhua 321004, China}
\hypersetup{pdfauthor={Shu He and Liwei Duan}}

\begin{abstract}
In the boundary time crystal of cooperative resonance fluorescence, the mean-field motion of $N$ driven, collectively decaying atoms is a continuous family of orbits with one period but different emission rates. Quantum noise selects no orbit; the atoms diffuse slowly across the family. The Fano factor of bare fluorescence thus grows as $N^2$, with a closed-form coefficient, and a clock counting these photons loses long-time precision as $1/N$. The period survives: in simulations up to $N=1600$, the click order at two detectors tracks the phase, giving ticks whose precision grows roughly as $N$.
\end{abstract}

\maketitle

Degenerate classical solutions are a recurring theme of many-body physics, and quantum fluctuations usually decide among them. In equilibrium, fluctuations select particular states and give the would-be soft modes a gap, as in order by disorder in frustrated magnets~\cite{Villain1980,Henley1989,Rau2018} and in cavity quantum electrodynamics~\cite{Schuler2020}. Driven-dissipative light--matter systems~\cite{Kirton2019,Hwang2018} raise the question for motion rather than for states. A clean example is cooperative resonance fluorescence: $N$ two-level atoms driven by a coherent field and decaying collectively through a common radiation channel~\cite{Drummond1978,Carmichael1980,Puri1979}. Above a critical drive, the mean-field Bloch vector neither comes to rest nor settles onto a single limit cycle: every member of a continuous family of closed orbits is a solution~\cite{Drummond1978,Cabot2023}. These persistent oscillations make the system a boundary time crystal~\cite{Iemini2018}; the nonequilibrium superradiant transition has been observed in a driven atomic cloud~\cite{Ferioli2023}, and continuous time crystals in atom--cavity and electron--nuclear spin systems~\cite{Kongkhambut2022,Greilich2024}.

\begin{figure*}[!htbp]
\centering
\includegraphics[width=\textwidth]{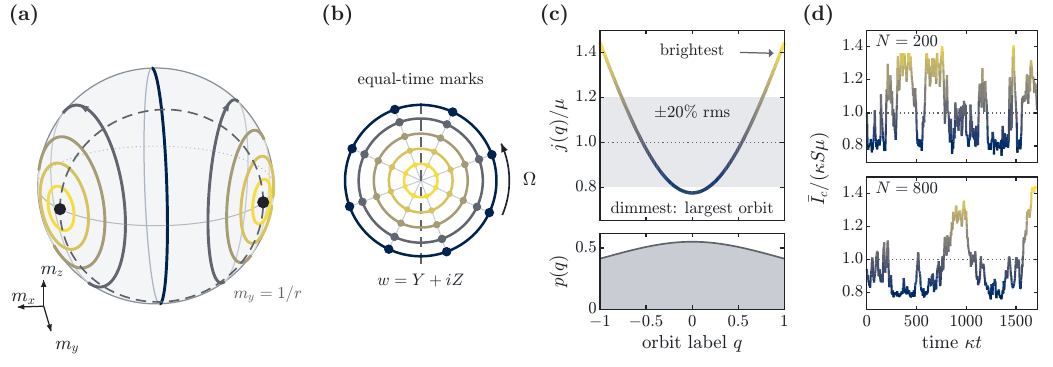}
\caption{\textbf{Degenerate orbit family of cooperative resonance fluorescence} ($r=2$). (a)~Mean-field orbits on the Bloch sphere, colored by their emission rate $j(q)$ [Eq.~\eqref{eq:brightness}] from dim (blue) to bright (yellow). The orbits shrink to two neutral centers at $q=\pm1$ (black dots); arrows, sense of motion; dashed, the plane $m_y=1/r$. (b)~In the variables of Eq.~\eqref{eq:qw} the motion is a rigid rotation at $\Omega=\kappa\sqrt{r^2-1}$: marks placed at equal times on all orbits stay on common spokes; dashed, ${\rm Re}\,w=0$. (c)~Orbit-averaged emission rate, normalized by its stationary mean $\mu$, and stationary orbit density $p(q)$ [Eq.~\eqref{eq:density}]; band, the $20\%$ rms spread. (d)~Wandering across the family along simulated quantum trajectories: conditional emission rate averaged over one period, normalized by $\kappa S\mu$, for $N=200$ and $800$. Its correlation time grows in proportion to $N$.}
\label{fig:orbits}
\end{figure*}

Do quantum fluctuations pin such a family, as they pin degenerate manifolds in equilibrium, and what does the emitted light reveal about it? Quantum trajectories show that the oscillation patterns coexist and that the system keeps jumping among them~\cite{Cabot2023}. Here we derive this motion and follow it into the light. Drift and diffusion across the family are both of order $1/N$, so no orbit is selected, and the atoms wander among the orbits on a time proportional to $N$ [Fig.~\ref{fig:orbits}]. Because the orbits share one period but emit at different rates, the photon flux remembers the current orbit, and the photon-number Fano factor grows as $N^2$; large count fluctuations in this regime were noted in a study of parameter estimation~\cite{Cabot2024}. Counting these photons has been proposed as a clock~\cite{Viotti2026}; we find that its long-time precision falls as $1/N$. The period, by contrast, survives the wandering and can be read from the same photons, through the order of clicks at two detectors.

\textit{Model and orbit family.}---We consider $N=2S$ atoms in the permutation-symmetric sector of total spin $S$, driven at Rabi frequency $\omega$ and decaying collectively at rate $\kappa$. With $\mathcal D[L]\rho=L\rho L^\dagger-\{L^\dagger L,\rho\}/2$ and $r=\omega/\kappa$, the master equation $\dot\rho=\mathcal L\rho$ reads
\begin{equation}
\mathcal L\rho=-i\omega[S_x,\rho]+\frac{\kappa}{S}\mathcal D[S_-]\rho=\frac{\kappa}{S}\mathcal D[A]\rho,
\label{eq:model}
\end{equation}
with $A=S_-+irS$. The two forms describe the same dynamics but different detection records. Direct photodetection of the fluorescence registers jumps of $S_-$; mixing the output with a coherent reference field before detection registers jumps of $A$. The latter are the counts of Ref.~\cite{Viotti2026}, which also considers a thermal version, $\mathcal L_T\rho=(\kappa/S)\{(\bar n+1)\mathcal D[A]\rho+\bar n\,\mathcal D[A^\dagger]\rho\}$ with $\bar n=1/(e^{\beta\omega_C}-1)$, in which $r$ is called $\lambda$; we take $\beta\omega_C=2$ as in Ref.~\cite{Viotti2026}. For $r<1$ the mean-field Bloch vector $\bm m=\langle\bm S\rangle/S$ relaxes to a fixed point; for $r>1$ it oscillates~\cite{Drummond1978,Carmichael1980}. We take $r=2$ unless stated otherwise.

Above threshold the mean-field equations conserve a hidden quantity~\cite{Iemini2018,Cabot2023}. With $a=\sqrt{r^2-1}$, the variables
\begin{equation}
q=\frac{a\,m_x}{r-m_y},\qquad w=\frac{r\,m_y-1+ia\,m_z}{r-m_y}
\label{eq:qw}
\end{equation}
obey $\dot q=0$ and $\dot w=i\Omega w$ with $\Omega=\kappa a$: $q$ labels the orbit, and $\arg w$ the position on it, which advances uniformly. Each orbit is the circle cut from the Bloch sphere by the plane $a m_x+q m_y=qr$. The circles form a coaxial family that shrinks to two neutral centers at $q=\pm1$ [Fig.~\ref{fig:orbits}(a)], and all of them have the same period $2\pi/\Omega$ [Fig.~\ref{fig:orbits}(b)]. The family is thus degenerate in frequency but not in brightness. Averaged over one period, the emission rate in units of $\kappa S$ is
\begin{equation}
j(q)=r^2-\frac{r a^2}{\sqrt{a^2+q^2}} ,
\label{eq:brightness}
\end{equation}
lowest on the largest orbit ($q=0$) and highest near the neutral centers [Fig.~\ref{fig:orbits}(c)], where, as we show below, the phase is hardest to read.

\textit{No selection, slow wandering.}---At mean field the orbit label $q$ is conserved, a soft coordinate of the family; quantum fluctuations act on it at order $1/N$. Averaged over the fast rotation, they produce a slow drift and diffusion of $q$, generated by $(\kappa/S)\mathscr G$ with $\mathscr G$ given in End Matter~A. Two questions matter for the light: how widely the atoms spread over the family, and how long they remember their orbit. Drift and diffusion are of the same order, so their ratio, and with it the stationary density
\begin{equation}
p(q)=\frac{a}{2\theta\,(a^2+q^2)},\qquad \theta=\arctan(1/a),
\label{eq:density}
\end{equation}
does not depend on $N$. No orbit is selected as $N\to\infty$, in contrast to order by disorder, and the brightness varies by $20\%$ rms about its mean $\mu=\int p\,j\,dq$. The memory is set by the relaxation rates $\lambda_k\kappa/S$, with $\lambda_k$ the eigenvalues of $-\mathscr G$; they appear in the exact Liouvillian spectrum as real eigenvalues that close as $1/N$~\cite{SM}. The soft mode thus stays soft, whereas in equilibrium zero-point fluctuations would give it a gap~\cite{Rau2018}. The brightness is even in $q$, so the slowest mode, which is odd, is invisible in the light; the brightness relaxes with the slowest even mode, $\lambda_1\approx1.75$ at $r=2$. Simulated quantum trajectories show the wandering directly [Fig.~\ref{fig:orbits}(d)].

\begin{figure}[!htbp]
\centering
\includegraphics[width=\columnwidth]{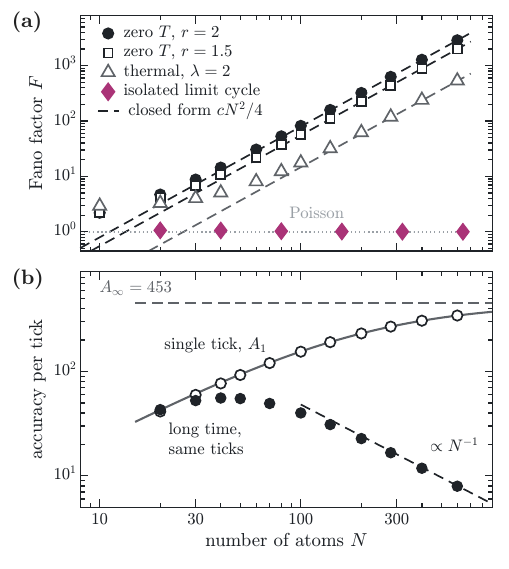}
\caption{\textbf{Photon counting on the orbit family.} (a)~Fano factor from the exact counting statistics: bare emission at zero temperature for $r=2$ and $1.5$, and emission counts in the thermal model ($\lambda=2$). Dashed lines: closed-form asymptotes (End Matter~A). Diamonds: a collective spin with an isolated limit cycle~\cite{Dutta2025}, at the loss strength where $F\to1$; for other values $F$ stays of order one~\cite{SM}. (b)~Accuracy per tick for ticks of $M=14.16\,S$ photons in the thermal model, the threshold of Ref.~\cite{Viotti2026} at $N=100$: single-tick accuracy $A_1$ (open circles) and long-time accuracy per tick $M/F$ (filled circles). Solid line, $A_\infty S/(S+S_*)$ with $A_\infty=453$ (dashed) and $S_*=96$, from optimized thresholds~\cite{SM}; dashed line at right, closed-form asymptote of $M/F$.}
\label{fig:counts}
\end{figure}

\textit{Photon statistics.}---The photon flux follows the brightness of the current orbit, with a contrast of order $\kappa S$ that persists for a time of order $S/\kappa$. The variance of the number $K(t)$ of emitted photons then grows as $(\kappa S)^2(S/\kappa)\,t=\kappa S^3t$, and the Fano factor $F=\lim_{t\to\infty}{\rm Var}\,K(t)/\langle K(t)\rangle=1+Q$, with $Q$ Mandel's parameter~\cite{Mandel1979}, grows quadratically,
\begin{equation}
F\simeq c(r)\,S^2=c(r)\,N^2/4 .
\label{eq:fano}
\end{equation}
The coefficient follows in closed form from the slow dynamics (End Matter~A); $c(2)=0.0318$. Exact counting statistics, obtained from the tilted Liouvillian~\cite{Landi2024}, approach this law in both models [Fig.~\ref{fig:counts}(a)]. The noise is a property of the degenerate family rather than of oscillation as such: for a collective spin with an isolated, attracting limit cycle~\cite{Dutta2025}, the Fano factor stays of order one for all $N$~\cite{SM}. It affects any detection scheme whose orbit-averaged signal depends on $q$, including homodyne detection of a fixed quadrature (End Matter~A).

\begin{figure*}[!htbp]
\centering
\includegraphics[width=\textwidth]{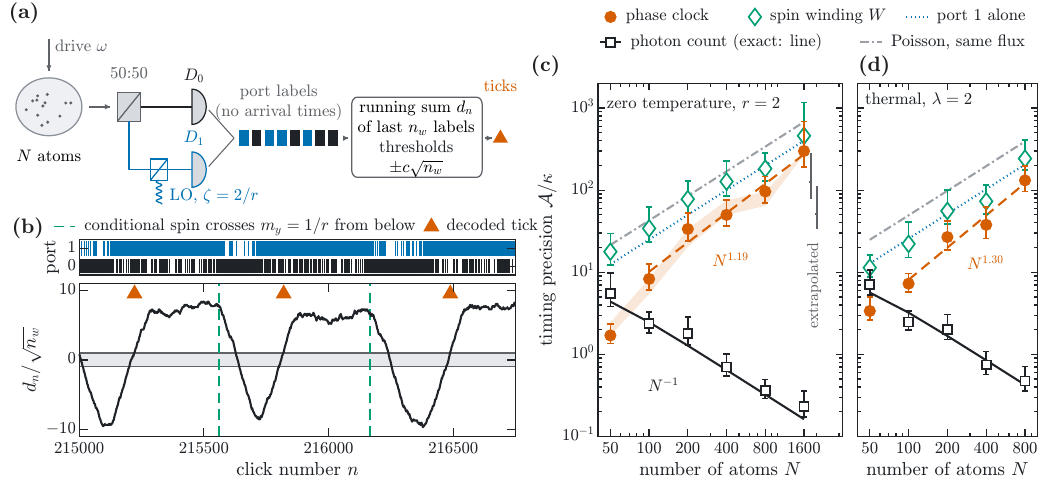}
\caption{\textbf{Reading the phase from the order of clicks.} (a)~Two-port detection, with port~1 displaced by a local oscillator, $\zeta=2/r$; the decoder receives only the sequence of port labels. (b)~A stretch of a record at $N=400$: port labels, normalized running sum $d_n/\sqrt{n_w}$ with the thresholds $\pm1$ (band), and decoded ticks, versus click number. The decoder never sees the conditional spin crossings (dashed). (c)~Timing precision of clocks built from the same photon records at zero temperature ($r=2$). Band, phase clock with windows $n_w\pm1$; dashed, fit for $N\ge100$; $W$, winding of the conditional spin phase; photon count, the total $N_0+N_1$ of the two ports; Poisson, a Poisson counter at the same mean flux; dotted, exact precision of port~1 alone, which here counts drive photons. Bars beyond $N=1600$: extrapolations from $N\le800$ assuming power-law growth (left) or a plateau (right). (d)~Same for the thermal model ($\lambda=2$), with absorption undetected. Error bars, 95\% bootstrap intervals.}
\label{fig:phase}
\end{figure*}

\textit{Counting clock.}---Counting photons gives a clock: $\hat t=K(t)/J$, with $J$ the mean photon flux, estimates the elapsed time. As is usual for autonomous clocks~\cite{Erker2017,Woods2022}, we quantify its long-time precision by $\mathcal A=\lim_{t\to\infty}t/{\rm Var}\,\hat t=J/F$, so that the relative timing error after a time $t$ is $(\mathcal At)^{-1/2}$. With $J\simeq\kappa\mu S$, Eq.~\eqref{eq:fano} gives $\mathcal A\simeq\kappa\mu/[c(r)S]$: the larger the crystal, the worse its photon-counting clock.

This long-time precision differs from the accuracy of a single tick, the measure studied in Ref.~\cite{Viotti2026}. If a tick is registered after $M$ photons, its accuracy, $A_1=\langle T\rangle^2/{\rm Var}\,T$, is blind to correlations between successive ticks, whereas the long-time accuracy per tick of the same ticks is $\mathcal A\langle T\rangle=M/F$. In the thermal model [Fig.~\ref{fig:counts}(b)], $A_1$ rises from $231$ to $307$ between $N=200$ and $400$, whereas $M/F$ falls from $22.8$ to $11.8$, and $A_1$ itself approaches a finite limit, $A_1\simeq A_\infty S/(S+S_*)$ with a semiclassical estimate $A_\infty=453$ and $S_*\simeq96$~\cite{SM}. Successive ticks come from nearly the same orbit, early again and again on a bright orbit and late on a dim one, over a number of ticks proportional to $N$, and these correlations dominate the accumulated error. Unlike the tick correlations that can improve a clock~\cite{Meier2026}, they degrade it here. A prepared state observed for a fixed short time behaves differently again~\cite{Yunoki2026}; the two limits do not conflict~\cite{SM}.

\textit{Reading the phase.}---The wandering leaves one quantity untouched. All orbits share the period $2\pi/\Omega$, so the phase $\phi=\arg w$ of the collective spin oscillation advances at the rate $\Omega$ on every orbit. To read it, we split the emitted field into two equal ports and displace port~1 by a local oscillator phase-locked to the drive; this changes the detection record but not Eq.~\eqref{eq:model} [Fig.~\ref{fig:phase}(a)]. With port operators $L_0=\sqrt{\kappa/(2S)}\,S_-$ and $L_1=\sqrt{\kappa/(2S)}\,(S_-+i\zeta S)$, the intensity difference is
\begin{equation}
L_1^\dagger L_1-L_0^\dagger L_0=\kappa\zeta\Big(\frac{\zeta S}{2}-S_y\Big).
\label{eq:ports}
\end{equation}
For $\zeta=2/r$ it changes sign exactly when the Bloch vector crosses the plane $m_y=1/r$, the line ${\rm Re}\,w=0$. Every orbit crosses this plane twice per period, once in each direction [Fig.~\ref{fig:orbits}(a,b)], so the port that clicks more often changes twice per cycle, whatever the orbit. Unlike phase-sensitive homodyne clocks~\cite{Singh2025}, the readout below uses only the order of photon labels.

\begin{figure*}[!htbp]
\centering
\includegraphics[width=\textwidth]{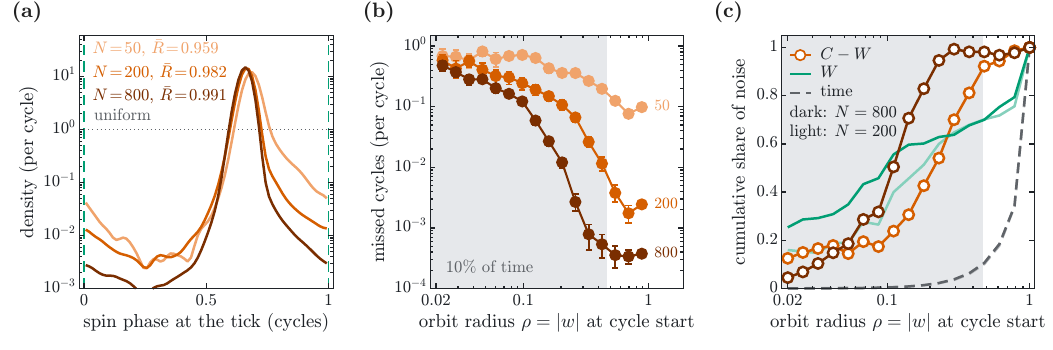}
\caption{\textbf{Where the phase clock loses time} (zero temperature, $r=2$). (a)~Distribution of the conditional spin phase at the decoded ticks, measured from the crossing of $m_y=1/r$ from below [dashed lines in Fig.~\ref{fig:phase}(b)], estimated with a von Mises kernel of bandwidth $1/48$ cycle. (b)~Probability that a cycle is missed, versus the radius $\rho=|w|$ of the conditional first moments at the start of the cycle. (c)~Cumulative share of the long-time timing variance carried by cycles that start below radius $\rho$, for the decoder error $C-W$ and the spin winding $W$; dashed, fraction of time. The shares partition each variance exactly, including covariances, and need not grow monotonically. Shading in (b,c), the smallest orbits, which occupy $10\%$ of the time at $N=800$.}
\label{fig:loss}
\end{figure*}

A simple decoder turns these labels into ticks. It keeps the running sum $d_n$ of the last $n_w$ labels ($+1$ for port~1, $-1$ for port~0), arms when $d_n<-c\sqrt{n_w}$, and fires a tick at the next $d_n>c\sqrt{n_w}$ [Fig.~\ref{fig:phase}(b)]; it thus registers the crossings in one direction, with $m_y$ falling through $1/r$. It uses neither arrival times nor any knowledge of the state, and $n_w$ is a fixed fraction $c_w$ of the clicks per period. We chose $c_w=1/4$ and $c=1$ at $N=50$ and $100$ and used them at every size (End Matter~C).

On identical photon records the two readings scale in opposite directions [Fig.~\ref{fig:phase}(c)]. For $N\ge100$ the precision of the phase clock grows as $N^{1.19}$ (95\% interval $[1.00,1.47]$), whereas that of photon counting falls as $N^{-0.90}$ (exact: $N^{-1}$). At $N=1600$ the precision of the phase clock exceeds that of photon counting on the same records by a factor of about $1300$ ($1800$ against the exact counting precision), and lies well above a plateau extrapolated from $N\le800$. The thermal model, in which absorption goes undetected, behaves the same way, with exponent $1.30$ $[1.11,1.51]$ [Fig.~\ref{fig:phase}(d)]. The growth persists for integer changes of the window and, over $N=100$--$400$, at detection efficiencies of $0.8$ and $0.5$~\cite{SM}.

Precision that grows with $N$ is not by itself evidence of a phase readout. A single port displaced by $\zeta=r/2$ counts drive photons: with $B=S_-+irS/2$, $(\kappa/S)B^\dagger B=\kappa r^2S/4-\mathcal L^\dagger S_z$, and by conservation of excitations this count differs from the Poissonian drive only by the bounded change of $S_z$~\cite{Cilluffo2019}. The displacement $\zeta=2/r$, which places the sign change at the plane $m_y=1/r$, equals $r/2$ only at $r=2$. There port~1 itself counts drive photons; its precision, $\kappa S/2$, grows linearly with $N$ [dotted line in Fig.~\ref{fig:phase}(c)], yet it measures the laser rather than the atoms. Two checks show that the phase clock does not borrow this flux. Shuffling the labels within each record keeps the click times and port totals, and with them any count of drive photons, but leaves the decoder with almost no ticks at large $N$ ($9$ instead of $241\,952$ in the first 64 records at $N=1600$). At $r=1.5$, where neither port counts drive photons, the phase clock still gains precision as $N^{1.01}$~\cite{SM}.

\textit{What the ticks register.}---Propagating the conditional state along each photon record gives the actual phase of the spin oscillation. At the decoded ticks this phase is sharply peaked, about $0.66$ cycles after the crossing of $m_y=1/r$ from below, that is, $0.16$ cycles in phase beyond the crossing from above, which the decoder registers [Fig.~\ref{fig:loss}(a)]. Its mean resultant length rises from $\bar R=0.959$ at $N=50$ to $0.991$ at $N=800$. The ticks thus follow the spin's cycles at a fixed phase, and the remaining errors are discrete: missed or extra cycles.

\textit{Where the phase clock loses time.}---Missed cycles occur almost only on the smallest orbits [Fig.~\ref{fig:loss}(b)]. We measure the orbit by the radius $\rho=|w|$ of the conditional first moments, which on a mean-field orbit equals $\sqrt{1-q^2}$. Cycles that start at $\rho\le0.31$ account for $85\%$ of the misses at $N=800$ but only $4\%$ of the time. There the label contrast of Eq.~\eqref{eq:ports}, proportional to $\rho$, is small, and the phase itself is poorly defined; these are also the brightest orbits. The winding number $W$ of the spin phase, an idealized reference computed from the full conditional state, loses time in the same place [Fig.~\ref{fig:loss}(c)]. When the long-time variance is split according to $\rho$ at the start of each cycle, with all covariances kept~\cite{SM}, the $10\%$ of time spent on the smallest orbits carries $98\%$ of the variance of the decoder error $C-W$, with $C$ the decoded count, and $70\%$ of that of $W$ at $N=800$. The phase clock reaches about half the precision of $W$ (ratio $0.53$ $[0.33,0.85]$), and the long-time fluctuations of the two are only weakly correlated (coefficient $0.37$). Because this ratio still grows with $N$, while $W$ gains precision about linearly, the fitted exponents in Fig.~\ref{fig:phase}(c,d) exceed one~\cite{SM}.

\textit{Discussion.}---A single fluorescence record thus carries two kinds of information. Its total count accumulates the slow brightness noise of the wandering, whereas the order of clicks at two ports marks directed crossings of the spin and loses contrast only where the oscillation amplitude vanishes. Isochrony makes the second reading possible: had the frequency depended on $q$, the same slow diffusion would have dephased the phase as well. These results assume strictly collective dynamics. Local dephasing at rate $\gamma_\phi$ breaks the conservation of $\bm S^2$, $\mathcal L^\dagger\bm S^2=-4\gamma_\phi(S_x^2+S_y^2-N/2)$, so in the true steady state the bare emission rate is at most $2\kappa$ for any $\gamma_\phi>0$, rather than of order $\kappa N$ (End Matter~B). The collective picture therefore holds only until local dephasing has depleted the maximal-spin sector. In equilibrium, fluctuations lift a degeneracy and open a gap~\cite{Villain1980,Henley1989,Rau2018}; here, at order $1/N$, they leave the family unpinned and make it visible in the statistics of the emitted light. Whether other open light--matter systems, such as coupled cavities, superradiant lattices~\cite{ZhaoHwang2022}, or driven-dissipative versions of frustrated Rabi triangles~\cite{HeDuan2026Triangle}, host such families and leave them unpinned is open; where they do and the orbits are isochronous, the light should show the same separation between brightness noise and phase.

\begin{acknowledgments}
L.D. acknowledges support from the National Natural Science Foundation of China (Grant No.~12305032) and the Zhejiang Provincial Natural Science Foundation of China (Grant No.~LQ23A050003).

\textit{AI use disclosure.}---OpenAI's ChatGPT and Codex assisted with language editing, checking derivations, optimizing numerical algorithms, and packaging the data for Zenodo. The authors verified the results and take responsibility for the manuscript.
\end{acknowledgments}

\bibliography{refs}

\section*{End Matter}
\setcounter{equation}{0}
\renewcommand{\theequation}{E\arabic{equation}}
\renewcommand{\theHequation}{E.\arabic{equation}}

\textit{A. Slow orbit dynamics and readout dependence.}---Averaging the fast rotation at fixed $q$ gives, to leading order in $1/S$, a reversible diffusion of $q$ generated by $[\kappa(2\bar n+1)/S]\,\mathscr G$, with
\begin{equation}
\mathscr Gf=\frac{1}{p}\big(D\,p\,f'\big)',\qquad D(q)=\frac{(a^2+q^2)(1-q^2)}{4a^2}.
\label{eq:generator}
\end{equation}
Here $\bar n=0$ at zero temperature and $2\bar n+1=\coth(\beta\omega_C/2)$ in the thermal model. The ends $q=\pm1$ are natural boundaries, and $p(q)$ of Eq.~\eqref{eq:density} is the stationary density. For a readout whose orbit-averaged current is $j_R(q)$, with mean $\mu_R=\int p\,j_R\,dq$, the zero-frequency noise follows from $\mathscr Gg=-(j_R-\mu_R)$ with $\int p\,g\,dq=0$. Integrating once gives $D\,p\,g'=-I_R$, with $I_R(q)=\int_{-1}^{q}p(u)\,[j_R(u)-\mu_R]\,du$, and
\begin{equation}
\frac{F_R}{S^2}\to\frac{2K_R}{(2\bar n+1)\,\mu_R},\qquad K_R=\int_{-1}^{1}\frac{I_R^2}{D\,p}\,dq\ \ge 0 .
\label{eq:criterion}
\end{equation}
$K_R$ vanishes only for an orbit-independent response. For bare emission, $j_R=j$, this gives
\begin{equation}
c(r)=\frac{8a^3}{\theta\mu}\int_0^1\frac{\big[\arctan(q/a)/\theta-rq/\sqrt{a^2+q^2}\big]^2}{1-q^2}\,dq ,
\label{eq:coefficient}
\end{equation}
with $c(1.5)=0.022138$, $c(2)=0.031842$, and $c(\infty)=2/45$. For the emission counts of the thermal model, Eq.~\eqref{eq:criterion} gives $F/S^2\to(\bar n+1)\,c\,\mu\,\theta/[(2\bar n+1)a]$, which equals $0.00587$ at $\lambda=2$, $\beta\omega_C=2$. The slowest even mode of $\mathscr G$ carries more than $99.9\%$ of the zero-frequency brightness noise. For counts of the displaced jump $S_-+i\zeta S$, the orbit-averaged rate is $j_\zeta=(1-2\zeta/r)\,j+\zeta^2$, so the leading term vanishes only at $\zeta=r/2$~\cite{SM}. A fixed homodyne quadrature, integrated over time, also has an orbit-dependent mean and therefore $K_R>0$. These are formal phase-averaged asymptotics. We checked them against exact Liouvillian spectra and counting statistics up to $N=600$ [Fig.~\ref{fig:counts}(a)]; a uniform $1/S$ expansion that also controls the boundary layers near $q=\pm1$, where the orbits shrink to points, remains an open problem.

\textit{B. Operator identities.}---Equation~\eqref{eq:ports} follows by expanding the two displaced quadratic forms. For $B=S_-+irS/2$, expanding $(\kappa/S)B^\dagger B$ and $\mathcal L^\dagger S_z$ gives $(\kappa/S)B^\dagger B=\kappa r^2S/4-\mathcal L^\dagger S_z$. The current of the half-displaced count is therefore a constant drive flux minus the time derivative of the bounded operator $S_z$, which cannot contribute to a long-time current. Since excitations are conserved, the count is the transmitted drive up to this bounded change, and all its long-time cumulants are Poissonian ($|F-1|<10^{-13}$ for port~1 at $r=2$). The weighted two-port count $X=(2-r^2/2)N_0+(r^2/2)N_1$ has the intensity $\kappa S-\mathcal L^\dagger S_z$. Its mean current is therefore independent of the orbit, and the leading brightness noise cancels; its higher cumulants are not Poissonian in general ($F_X=1.044$ at $r=1.5$, $S=32$). At $r=2$, $X=2N_1$ and $\mathcal A_X=\kappa S/2$. At $r=\sqrt2$ the two weights are equal: mixing the two equal-weight channels unitarily gives the drive counter $B$ and a constant jump operator, so the plain sum of the two port counts has Poissonian long-time statistics, with $F=1$ and $\mathcal A=\kappa S$ (checked exactly for $N\le128$). Whether a photon count inherits the slow noise thus depends on whether its orbit-averaged rate resolves the orbits. For local dephasing, $\gamma_\phi\sum_i\mathcal D[\sigma_i^z]$, conjugation by $\sigma_i^z$ reverses the transverse components of spin $i$, and summing the cross terms gives $\sum_i\mathcal D^\dagger[\sigma_i^z]\,\bm S^2=-4(S_x^2+S_y^2-N/2)$, which we also verified in the full $2^N$-dimensional space for $N\le6$. The collective generator conserves $\bm S^2$, so a stationary state with $\gamma_\phi>0$ has $\langle S_x^2+S_y^2\rangle=N/2$ and hence $\langle S_+S_-\rangle=N/2+\langle S_z\rangle\le N$: the bare emission rate $(\kappa/S)\langle S_+S_-\rangle$ is at most $2\kappa$. A displaced detector adds a coherent flux, to which this bound does not apply.

\textit{C. Decoder and extrapolation.}---For labels $\ell_n\in\{0,1\}$, $d_n=\sum_{k=n-n_w+1}^{n}(2\ell_k-1)$. The decoder starts unarmed, arms at the first $d_n<-c\sqrt{n_w}$, emits one tick at the next $d_n>c\sqrt{n_w}$, and disarms. The window is $n_w=c_w\,\nu_{\rm click}T$, where $\nu_{\rm click}$ is the click rate at the given $N$ and $T=2\pi/\Omega$; the labels are not mean-subtracted. Of ten settings ($c_w\in\{1/8,1/4,1/2\}$ and $c\in\{1,2,3\}$, plus a debounced variant), we chose $c_w=1/4$, $c=1$ by comparing their precision at $N=50$ and $100$ only, and used them at every size. The precision is $\mathcal A=\nu^2/V_C$, where $\nu$ is the tick frequency and $V_C=d\,{\rm Var}\,C/dt$ is obtained from an affine fit of ${\rm Var}\,C(t)$ over $[15.25,30]\,\tau_{\rm slow}$. The click rate, the tick frequency, and the choice of setting come from trajectories separate from those used for the results~\cite{SM}. Before simulating $N=1600$, we extrapolated the precision from $N\le800$ in three forms~\cite{SM}; the two shown in Fig.~\ref{fig:phase}(c) give $127$ $[78,281]\,\kappa$ for a power law and $52$ $[33,111]\,\kappa$ for a plateau. The simulation gives $300$ $[193,688]\,\kappa$, which excludes the plateau. A decoder that subtracts the mean label imbalance and scales its thresholds with $n_w$ fails: cycles on small orbits, whose label contrast is weak, then miss the thresholds, and the precision falls as $N^{-1.90}$~\cite{SM}.

\textit{D. Numerical methods.}---Exact results use the $(2S+1)$-dimensional symmetric space. Stationary states and zero-frequency cumulants follow from sparse LU solves with a trace constraint: $J={\rm Tr}\,\mathcal J_1\rho_{\rm ss}$ and the variance growth rate $V={\rm Tr}\,\mathcal J_2\rho_{\rm ss}+2\,{\rm Tr}\,\mathcal J_1y$, so that $F=V/J$, where $\mathcal L y=-(\mathcal J_1-J)\rho_{\rm ss}$ with ${\rm Tr}\,y=0$, and $\mathcal J_{1,2}$ are the first and second derivatives of the tilted generator. At zero temperature, the exact stationary state $\rho_{\rm ss}\propto(A^\dagger A)^{-1}$~\cite{Puri1979} serves as a check. Single-tick moments follow from the $M$-fold first-passage propagator applied to the stationary post-count state. Trajectories use the two-port jump unraveling with exact sampling of waiting times, start from the stationary state, and run for $30\,\tau_{\rm slow}$, where $\tau_{\rm slow}$ is the inverse rate of the slowest even real Liouvillian eigenvalue at that size, which tends to $S/[\kappa(2\bar n+1)\lambda_1]$; their number at each size is listed in the SM~\cite{SM}. Along each record the conditional state is propagated and stored at every click and at 64 equally spaced times per period. In the thermal model the decoder uses only the detected emission labels, whereas the reference winding $W$ is computed from the full simulated trajectory, including the undetected absorption events. Intervals are $95\%$ percentile intervals from bootstrap resampling of whole trajectories. The data underlying the figures, the results of the exact calculations, and the simulation parameters and random seeds are available on Zenodo~\cite{HeDuan2026Data}.

\end{document}


\setcounter{secnumdepth}{1}
\title{Supplemental Material for ``Diffusion among degenerate orbits and phase timekeeping in a boundary time crystal''}
\hypersetup{pdftitle={Supplemental Material for Diffusion among degenerate orbits and phase timekeeping in a boundary time crystal}}
\author{Shu He}
\email{heshu1987@foxmail.com}
\affiliation{Department of Physics and Electronic Engineering, Sichuan Normal University, Chengdu 610066, China}
\author{Liwei Duan}
\email{duanlw@gmail.com}
\affiliation{Department of Physics, Zhejiang Normal University, Jinhua 321004, China}
\hypersetup{pdfauthor={Shu He and Liwei Duan}}

\maketitle

Throughout, $\kappa=1$ sets the unit of rates, $N=2S$ is the number of atoms, and $r=\omega/\kappa$. ``Zero temperature'' refers to Eq.~(1) of the main text; the ``thermal model'' is the model of Ref.~\cite{Viotti2026},
\begin{equation}
\mathcal L_T\rho=\frac{\kappa}{S}\Big\{(\bar n+1)\,\mathcal D[A]\rho+\bar n\,\mathcal D[A^\dagger]\rho\Big\},\qquad A=S_-+i\lambda S,
\label{eq:thermal}
\end{equation}
with $\lambda\equiv r$ and $\bar n=1/(e^{\beta\omega_C}-1)$, $\beta\omega_C=2$. Unless stated otherwise, only emission ($A$) events are counted in the thermal model.

\section{Exact calculations and checks}

\textit{Liouvillian and counting statistics.}---All exact results use the permutation-symmetric space $\{|S,m\rangle\}$ of dimension $2S+1$. The Liouvillian is a sparse matrix of dimension $(2S+1)^2$ in the column-stacking convention ${\rm vec}(XYZ)=(Z^{T}\otimes X)\,{\rm vec}(Y)$. Counting a set of jump channels $L_i$ with weights $w_i$ introduces the tilted generator $\mathcal L(\chi)=\mathcal L+\sum_i(e^{i\chi w_i}-1)\mathcal J_i$, $\mathcal J_i\rho=L_i\rho L_i^\dagger$. With $\mathcal J_1=\sum_iw_i\mathcal J_i$ and $\mathcal J_2=\sum_iw_i^2\mathcal J_i$, the mean current and the long-time variance growth rate are
\begin{equation}
J={\rm Tr}\,\mathcal J_1\rho_{\rm ss},\qquad V={\rm Tr}\,\mathcal J_2\rho_{\rm ss}+2\,{\rm Tr}\,\mathcal J_1 y ,
\end{equation}
where $\mathcal L y=-(\mathcal J_1-J)\rho_{\rm ss}$ with ${\rm Tr}\,y=0$~\cite{Landi2024}. We solve both the stationary and the Poisson equation with one sparse LU factorization of $\mathcal L$, in which the row of the $(m,m)=(S,S)$ element is replaced by the trace condition. The Fano factor is $F=V/J$ and the clock precision is $\mathcal A=J/F$. At finite time,
\begin{equation}
{\rm Var}\,K(t)=t\,{\rm Tr}\,\mathcal J_2\rho_{\rm ss}+2\,{\rm Tr}\,\mathcal J_1\big[\,t\,y-z+e^{\mathcal Lt}z\,\big],
\end{equation}
with $\mathcal L z=-(y-{\rm Tr}\,y\,\rho_{\rm ss})$ and ${\rm Tr}\,z=0$. Bare emission at zero temperature counts $L=S_-$ with the drive $\omega S_x$ in the Hamiltonian; the counts of Ref.~\cite{Viotti2026} use the displaced operator $A$, which generates the same dynamics but a different record.

\textit{Isolated limit cycle.}---The control model of Fig.~2(a) of the main text is a collective spin of the kind studied in Ref.~\cite{Dutta2025}, whose mean-field motion is an isolated, attracting limit cycle: $H=-S_z$, collective gain $L_+=S_+/\sqrt S$, and nonlinear loss $L_-=\sqrt{\gamma/S^3}\,S_-S_z$ with $\gamma=3$, in units where the precession frequency is one, restricted to the invariant sector $0\le m\le S$. Both jump channels are counted, and $F$ follows from the expressions above for $S=10$ to $320$. Since $H$ commutes with $S_z$ and both jumps change $m$ by one, the populations obey a birth--death chain, which gives the same values. On the cycle, $z_*=m/S=1/\sqrt\gamma$; for large $S$ the total jump rate is $S\,h(z)$ with $h(z)=(1-z^2)(1+\gamma z^2)$, and $z$ relaxes at the rate $2z_*(\gamma-1)$. A linear-noise expansion about the cycle gives $F\to1+[(\gamma-3)/(\gamma-1)]^2$, independent of $S$: $F$ stays of order one for every $\gamma>1$ and tends to one only at $\gamma=3$, where $h'(z_*)=2z_*(\gamma-3)$ vanishes. At $S=320$ we find $F=2.039$, $1.003$, and $1.111$ for $\gamma=2$, $3$, and $4$, against the limits $2$, $1$, and $10/9$.

\textit{Single ticks.}---A tick is registered after $M$ counts. With $\mathcal L_0=\mathcal L-\mathcal J_1$ the no-count generator and $G(s)=\mathcal J_1(s-\mathcal L_0)^{-1}$, the Laplace transform of the tick interval is ${\rm Tr}\,G(s)^M\rho_*$, where $\rho_*=\mathcal J_1\rho_{\rm ss}/J$ is the stationary state just after a count. We propagate the vector together with its first and second derivatives in $s$, which requires one LU factorization of $\mathcal L_0$ and six solves per count. This gives $\langle T\rangle=M/J$ and $A_1=\langle T\rangle^2/{\rm Var}\,T$ for a tick in a stationary tick sequence.

\begin{figure}[!htbp]
\centering
\includegraphics[width=\columnwidth]{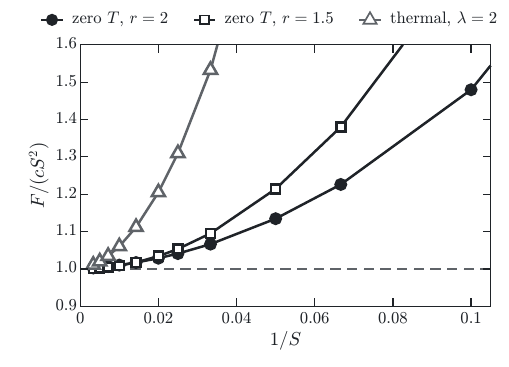}
\caption{Approach of the exact Fano factors to their closed-form limits, $F/(cS^2)\to1$, with $c=c(r)$ for bare emission at zero temperature ($r=2$, $1.5$) and $c=(\bar n+1)\,c(r)\,\mu\theta/[(2\bar n+1)a]=0.00587$ for emission counts in the thermal model ($\lambda=2$; End Matter~A). The corrections are of order $1/S$.}
\label{fig:coef}
\end{figure}

\textit{Checks.}---Table~\ref{tab:checks} lists the checks we made. The exact lines in Fig.~3(c,d) of the main text are computed in the same way, with the two-port jump operators of Sec.~\ref{sec:decoder}. Figure~\ref{fig:coef} shows how the exact Fano factors approach the closed-form coefficients.

\begin{table}[t]
\caption{Checks of the exact calculations (largest deviations unless stated). From top: stationary state against the numerical solution; the two forms of Eq.~(1) of the main text; numerical integration of the mean-field equations against the rigid rotation of Eq.~(2); Eq.~(E3) against the sum over the modes of $\mathscr G$; mean single-tick interval; single-tick accuracy in the thermal model, with the values of an independent implementation (sparse first-passage moment recursion) in parentheses; Fano factor of port~1, a drive counter at $r=2$.}
\label{tab:checks}
\begin{ruledtabular}
\begin{tabular}{lr}
Check & Result\\\hline
$\rho_{\rm ss}\propto(A^\dagger A)^{-1}$, $S\le50$ & $2\times10^{-15}$\\
Eq.~(1), two forms, $S\le50$ & $1\times10^{-14}$\\
Rigid rotation, 20 orbits & $4\times10^{-10}$\\
Eq.~(E3) vs modes of $\mathscr G$ & $2\times10^{-7}$ (rel.)\\
$\langle T\rangle J/M-1$, $S=25,50$ & $3\times10^{-14}$\\
$A_1$, $\lambda=1.5$, $S=25$, $M=257$ & $28.878$ $(28.88)$\\
$A_1$, $\lambda=2$, $S=50$, $M=708$ & $154.767$ $(154.77)$\\
$|F-1|$, port~1, $r=2$, $S\le100$ & $<10^{-13}$\\
\end{tabular}
\end{ruledtabular}
\end{table}

\section{Mean-field orbit family}\label{sec:mf}

With $\bm m=\langle\bm S\rangle/S$, the mean-field equations of Eq.~(1) are
\begin{align}
\dot m_x&=\kappa m_xm_z,\qquad \dot m_y=-\omega m_z+\kappa m_ym_z,\nonumber\\
\dot m_z&=\omega m_y-\kappa(m_x^2+m_y^2).
\end{align}
Since $m_x$ and $m_y-r$ obey the same equation, $\dot u=\kappa m_z u$, their ratio is conserved; this is $q$ of Eq.~(2) up to a factor~\cite{Iemini2018,Cabot2023}. Writing $w=Y+iZ$, the inverse map
\begin{equation}
m_x=\frac{aq}{r+Y},\qquad m_y=\frac{1+rY}{r+Y},\qquad m_z=\frac{aZ}{r+Y}
\end{equation}
satisfies $Y^2+Z^2+q^2=1$ on the unit sphere, and the denominator is positive for $r>1$. Substituting into the equations of motion gives $\dot w=i\Omega w$ with $\Omega=\kappa a$. Uniform angles in the $(Y,Z)$ plane are therefore uniform times, which is why equal-time marks stay on common spokes in Fig.~1(b). The plane $m_y=1/r$ is the line $Y=0$, crossed at the phases $\pm\pi/2$. The neutral centers, $q=\pm1$, are the points $(\pm a/r,1/r,0)$; they lie on this plane.

Averaging $\dot m_z$ over one period gives $r\,\overline{m_y}=\overline{m_x^2+m_y^2}$, the orbit-averaged emission rate in units of $\kappa S$. With $Y=\rho\cos\phi$ and $\rho^2=1-q^2$, the average $\overline{1/(r+Y)}=1/\sqrt{r^2-\rho^2}$ gives $\overline{m_y}=r-a^2/\sqrt{a^2+q^2}$ and hence Eq.~(3).

\section{Slow orbit dynamics and the Liouvillian spectrum}

\begin{figure}[!htbp]
\centering
\includegraphics[width=\columnwidth]{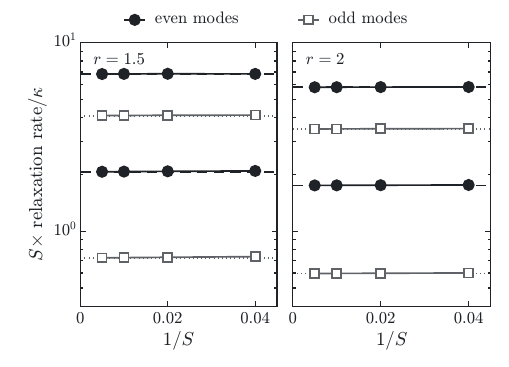}
\caption{Slow relaxation rates of the exact Liouvillian (symbols, decay rates scaled by $S/\kappa$) for $S=25$ to $200$, compared with the eigenvalues of $-\mathscr G$ (lines). Even modes (filled) couple to the brightness; odd modes (open) do not.}
\label{fig:spectrum}
\end{figure}

\textit{Derivation.}---Three ingredients fix the slow generator $\mathscr G$ of Eq.~(E1).

(i) \textit{Stationary density.} For $r>1$, $|\langle A\rangle|^2/S^2=m_x^2+(r-m_y)^2\ge(r-1)^2>0$ on the whole Bloch sphere. At large $S$ the exact stationary state $\rho_{\rm ss}\propto(A^\dagger A)^{-1}$~\cite{Puri1979} therefore corresponds to a density proportional to $[m_x^2+(r-m_y)^2]^{-1}$ on the sphere. The map from the unit vector $(q,Y,Z)$ to $\bm m$ in Sec.~\ref{sec:mf} is a Lorentz boost of the sphere, with $\cosh\eta=r/a$; it rescales lengths by $a/(r+Y)$ and areas by $a^2/(r+Y)^2$, while $m_x^2+(r-m_y)^2=a^2(a^2+q^2)/(r+Y)^2$. In the coordinates $(q,Y,Z)$ the density is thus proportional to $1/(a^2+q^2)$ and uniform in the angle around the $q$ axis. Since the area element of the unit sphere is $dq\,d\phi$, the density of $q$ is $p(q)=a/[2\theta(a^2+q^2)]$, Eq.~(4) of the main text.

(ii) \textit{Diffusion.} For a state localized near $\bm m$, with covariances of order $S$, the identity
\begin{equation}
\mathcal L^\dagger(XY)-X\,\mathcal L^\dagger Y-(\mathcal L^\dagger X)\,Y=\frac{\kappa}{S}\,[A^\dagger,X][Y,A]
\end{equation}
shows that the dissipator adds $(\kappa/S)\,{\rm Re}\,\langle[A^\dagger,S_i][S_j,A]\rangle$ per unit time to ${\rm Cov}(S_i,S_j)$. To leading order $[S_j,A]=S\,c_j$ with $\bm c=(m_z,\,-im_z,\,im_y-m_x)$. The mean-field flow conserves $q$, so its linearization leaves the variance of $q$ unchanged, and only this noise spreads it:
\begin{equation}
\frac{d}{dt}{\rm Var}\,q=\frac{\kappa}{S}\,\big|\nabla q\cdot\bm c\big|^2=\frac{\kappa}{S}\,\frac{(a^2+q^2)\,Z^2}{a^2}.
\end{equation}
The phase advances uniformly, so the average over one period is $\overline{Z^2}=(1-q^2)/2$, which gives $d\,{\rm Var}\,q/dt=2(\kappa/S)D(q)$ with $D(q)$ of Eq.~(E1). The bare form of Eq.~(1) gives the same $\bm c$, because a $c$-number displacement does not change commutators. In the thermal model the $A^\dagger$ channel adds noise of the same size with weight $\bar n$, so that $D$ is multiplied by $2\bar n+1$.

(iii) \textit{Drift.} A one-dimensional diffusion with zero flux at the natural boundaries $q=\pm1$, where $D$ vanishes, is fixed by $D$ and its stationary density; its generator is $p^{-1}(Dpf')'$, which is Eq.~(E1).

These steps give the leading order in $1/S$ for times long compared with one period. Near $q=\pm1$, where the orbits shrink to points, the averaging is not uniform (End Matter~A), and the check is the exact spectrum.

\textit{Spectrum.}---The phase-averaged generator $\mathscr G$ of Eq.~(E1) is self-adjoint with respect to $p(q)\,dq$ and separates into modes even and odd in $q$. Its eigenvalues are listed in Table~\ref{tab:modes}. Multiplied by $S/\kappa$, the slowest decay rates of the exact Liouvillian approach them with corrections of order $1/S$ [Fig.~\ref{fig:spectrum}]; at $S=200$ the exact values of $S\times$rate are $0.7225$, $2.0676$, $4.1045$, $6.8211$ ($r=1.5$) and $0.5967$, $1.7500$, $3.4873$, $5.8042$ ($r=2$). The brightness $j(q)$ is even, so only even modes enter the zero-frequency photon noise, and the slowest even mode carries almost all of it.

In the thermal model the slow generator is multiplied by $2\bar n+1=\coth(\beta\omega_C/2)$: at $\lambda=2$ and $S=100$, the slowest Liouvillian decay rates, times $S/[\kappa(2\bar n+1)]$, are $0.5973$, $1.7515$, $3.4899$, and $5.8081$, against $0.5961$, $1.7485$, $3.4843$, and $5.7994$ for $-\mathscr G$. For a count with weights $w_-$ and $w_+$ on emission and absorption events, Eq.~(E2) gives $F/S^2\to2RK/[(2\bar n+1)\mu_V]$ with $R=w_-(\bar n+1)+w_+\bar n$, $\mu_V=a/\theta$, and $K=c\mu/2$ the Dirichlet energy of bare emission. At $\lambda=1.5$ this predicts $0.009136$, $0.010373$, and $0.007900$ for emission, total, and net counts; exact calculations up to $S=300$ extrapolate to these values within $0.02\%$. At $\lambda=2$ the emission coefficient is $0.00587$ [Fig.~\ref{fig:coef}].

\begin{table}[t]
\caption{Lowest eigenvalues of $-\mathscr G$, labeled by parity, and the share $w_1$ of the zero-frequency brightness noise carried by the slowest even mode.}
\label{tab:modes}
\begin{ruledtabular}
\begin{tabular}{lccccc}
$r$ & odd & even & odd & even & $w_1$\\\hline
$1.5$ & $0.7207$ & $2.0631$ & $4.0960$ & $6.8077$ & $99.93\%$\\
$2$ & $0.5961$ & $1.7485$ & $3.4843$ & $5.7994$ & $99.98\%$\\
\end{tabular}
\end{ruledtabular}
\end{table}

\section{Dependence on the readout}

\begin{figure}[!htbp]
\centering
\includegraphics[width=\columnwidth]{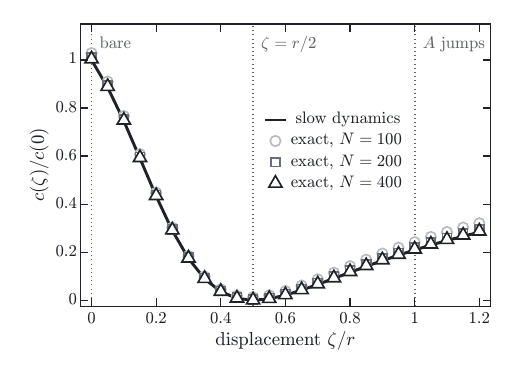}
\caption{Leading noise coefficient of displaced counting relative to bare counting, from the slow dynamics (line) and from exact counting statistics at $N=100$, $200$, $400$ (symbols), at $r=2$. The only zero is at $\zeta=r/2$, where the count measures drive photons.}
\label{fig:displaced}
\end{figure}

\textit{Displaced counting.}---Counting the jumps of $\sqrt{\kappa/S}\,(S_-+i\zeta S)$, with the drive reduced to $\kappa(r-\zeta)S_x$ so that the dynamics is unchanged, the orbit-averaged rate is $j_\zeta=(1-2\zeta/r)\,j+\zeta^2$. Equation~(E2) then gives
\begin{equation}
\frac{c(\zeta)}{c(0)}=\frac{\mu\,(1-2\zeta/r)^2}{(1-2\zeta/r)\,\mu+\zeta^2},
\end{equation}
which vanishes only at $\zeta=r/2$ and reduces to the coefficient of the $A$ jumps at $\zeta=r$ [Fig.~\ref{fig:displaced}]. Exact finite-size values approach this curve; at $\zeta=r/2$ the remaining noise is of lower order in $S$.

\textit{Drive counters.}---At $\zeta=r/2$ the count is the drive light transmitted through the atoms. Because excitations are conserved, it differs from the Poissonian photon number of the coherent drive only by the change of $S_z$, which is bounded (End Matter~B), so its long-time statistics are Poissonian; for port~1 at $r=2$ we find $|F-1|<10^{-13}$ (Table~\ref{tab:checks}). Its clicks are nevertheless correlated in time: at $r=2$ and $S=25$, $g^{(2)}(0)=2.137$, $g^{(2)}(T/2)=0.324$, and $g^{(2)}(T)=1.901$, with $T=2\pi/\Omega$. Such a counter improves with $N$ because its flux grows, not because it follows the atoms. The plain sum of the two ports at $r=\sqrt2$ (End Matter~B) behaves in the same way: $F=1$, but $g^{(2)}(0)=1.429$ at $S=32$.

\textit{Homodyne detection.}---The time integral of a fixed homodyne quadrature has an orbit-dependent mean, so Eq.~(E2) gives $K_R>0$. If the stationary mean current of the quadrature is nonzero, so that $\hat t=K/J$ can be calibrated, its precision falls as $1/N$ in the same way; a quadrature with zero mean current gives no such clock. This statement concerns an unfiltered integral and does not exclude phase-sensitive processing of the homodyne record~\cite{Singh2025}. Estimating the drive frequency from the full record, which improves faster with $N$~\cite{OConnor2025}, is a different task from keeping time with one fixed calibration.

\section{Timekeeping with photon counts}

\begin{figure}[!htbp]
\centering
\includegraphics[width=\columnwidth]{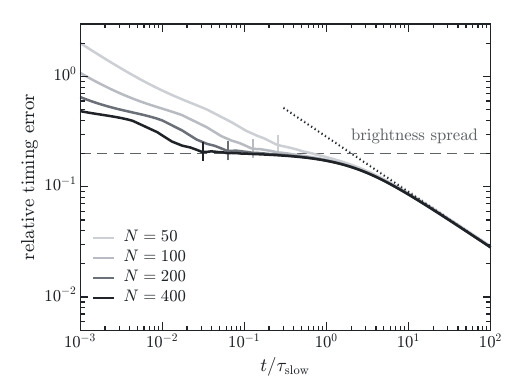}
\caption{Relative timing error of the photon-count clock (bare emission, $r=2$) started in the stationary state, from the exact finite-time variance. Short vertical marks: one period. Dashed: rms brightness spread. Dotted: long-time asymptotes $(\mathcal At)^{-1/2}$ with the exact $\mathcal A=J/F$.}
\label{fig:finite}
\end{figure}

\textit{Finite time.}---Figure~\ref{fig:finite} shows the relative timing error $\sqrt{{\rm Var}\,\hat t}/t$ of the photon-count clock started in the stationary state. After about one period the curves for all $N$ join a plateau at the rms brightness spread, $20\%$, and they decrease as $(\mathcal At)^{-1/2}$ only once $t$ exceeds $\tau_{\rm slow}=S/(\kappa\lambda_1)$, where $\lambda_1$ is the slowest even eigenvalue of $-\mathscr G$ (Table~\ref{tab:modes}): a stationary count cannot tell the orbit apart from the clock rate until the orbit has been forgotten. The plateau therefore needs $T\ll t\ll\tau_{\rm slow}$.

\begin{figure}[!htbp]
\centering
\includegraphics[width=\columnwidth]{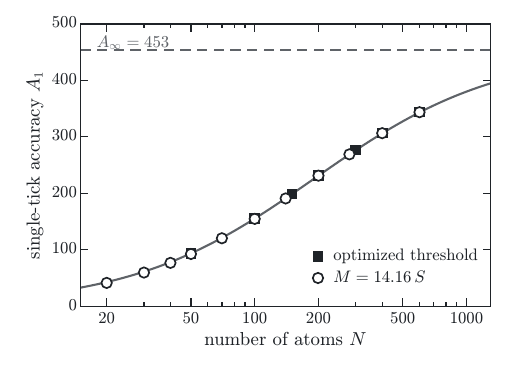}
\caption{Single-tick accuracy in the thermal model ($\lambda=2$) with optimized thresholds (squares) and with $M=14.16\,S$ (circles). Line: $A_\infty S/(S+S_*)$ with $S_*=95.9$; dashed: $A_\infty=453$.}
\label{fig:single}
\end{figure}

\textit{Single ticks.}---In the thermal model at $\lambda=2$, a semiclassical estimate gives the limit of the single-tick accuracy, $A_\infty=453.09$. With $A_\infty$ fixed at this value, $A_1=A_\infty S/(S+S_*)$ with $S_*=95.9$ describes the exact values with optimized thresholds within $0.3\%$ [Fig.~\ref{fig:single}]. Independently, a free fit of $1/A_1$ linear in $1/S$ to $N\le300$ gives $A_\infty=457.0$ and $S_*=97.3$ and predicts $N=400$ and $600$ within $0.4\%$. With the fixed resolution $M=14.16\,S$ used in Fig.~2(b) of the main text ($M=708$ at $S=50$, as in Ref.~\cite{Viotti2026}), the values lie within $1\%$ of the optimized ones. With the threshold at the first local minimum of ${\rm Var}\,T/\langle T\rangle$, as in Ref.~\cite{Viotti2026}, the limit is $452.53$. Our values are exact averages over the stationary state, so their variance includes the spread of brightness over the orbit family; the tick intervals in the data set of Ref.~\cite{Zenodo2026} have the same means, to a relative $7.6\times10^{-4}$ at equal count and threshold, and variances smaller by $0.8\%$ to $28\%$ for $S=25$ to $300$.

\begin{figure}[!htbp]
\centering
\includegraphics[width=\columnwidth]{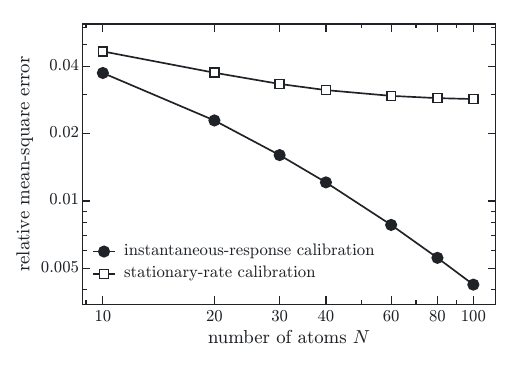}
\caption{Fixed-time protocol of Ref.~\cite{Yunoki2026}: relative mean-square error of the elapsed-time estimate with the instantaneous response (filled circles) and the stationary rate (open squares) used for calibration; exact moments.}
\label{fig:yunoki}
\end{figure}

\textit{Prepared state, fixed time.}---Yunoki and Hasegawa consider a spin coherent state prepared at the north pole and observed for a fixed duration~\cite{Yunoki2026}. At $\kappa\tau=10$, we estimate the elapsed time from the photon count $K$ using $\hat\tau = \tau + [K - \langle K\rangle]/\partial_\tau\langle K\rangle$ for instantaneous-response calibration and $\hat\tau=K/J$ for stationary-rate calibration. The plotted quantity is $\langle(\hat\tau - \tau)^2\rangle/\tau^2$. Our exact recalculation reproduces the enhancement they find at short times with the instantaneous-response calibration; with the stationary rate as calibration, the error decreases only weakly with $N$ [Fig.~\ref{fig:yunoki}]. This does not contradict the long-time result of the main text: the two protocols differ in preparation and in the order of limits.

\section{Phase clock: protocol and robustness}\label{sec:decoder}

\textit{Detection.}---At zero temperature the two ports are $L_0=\sqrt{\kappa/(2S)}\,S_-$ and $L_1=\sqrt{\kappa/(2S)}\,(S_-+i\zeta S)$ with $\zeta=2/r$; together with the Hamiltonian $\kappa(r-\zeta/2)S_x$ they reproduce Eq.~(1). Physically, port~1 is mixed with a local oscillator phase-locked to the drive. The mixing changes the detection record but not the dynamics, and the local oscillator carries no modulation at $\Omega$. In the thermal model the emission jump is split in the same way, with rate factor $\bar n+1$ and Hamiltonian $(\bar n+1)\kappa(r-\zeta/2)S_x$; absorption events are not detected. Trajectories are generated with the two-port jump unraveling. Waiting times are sampled exactly by inverting the survival probability, and initial pure states are drawn from an ensemble whose average is $\rho_{\rm ss}$. Table~\ref{tab:ntraj} lists the number of trajectories at each size.

\begin{table}[t]
\caption{Number of simulated trajectories used for the precision and the error budget, each of duration $30\,\tau_{\rm slow}$. The shifted windows and the thinned records of Fig.~\ref{fig:robust}(a,b) use the zero-temperature records at $r=2$; at $N=1600$, $W$ and the error budget use the first 64 records. The decoder setting, the click rates, and the tick frequencies were obtained from further trajectories (Sec.~\ref{sec:decoder}).}
\label{tab:ntraj}
\begin{ruledtabular}
\begin{tabular}{rccc}
$N$ & zero $T$, $r=2$ & thermal, $\lambda=2$ & zero $T$, $r=1.5$\\\hline
50 & 256 & 256 & 256\\
100 & 256 & 256 & 256\\
200 & 256 & 256 & 256\\
400 & 256 & 256 & 256\\
800 & 456 & 256 & 256\\
1600 & 256 & -- & --\\
\end{tabular}
\end{ruledtabular}
\end{table}

\textit{Decoder.}---The decoder is defined in End Matter~C. It uses strict inequalities and full windows, and the integer window is set by the click rate measured on separate trajectories. We compared the ten settings of Table~\ref{tab:settings} at $N=50$ and $100$ only. At each of these sizes, 128 trajectories gave the click rate and 128 others the precision of each setting at the end of the run; setting~4 ($c_w=1/4$, $c=1$) had the highest geometric mean of its precisions at the two sizes and is used throughout. The tick frequencies are also measured on separate trajectories, so the trajectories of Table~\ref{tab:ntraj} enter neither the choice of setting nor the calibration. Table~\ref{tab:extrap} lists three extrapolations of the precision to $N=1600$ from $N\le800$; the simulation gives $300$ $[193,688]\,\kappa$.

\begin{table}[t]
\caption{Decoder settings. With zero threshold ($c=0$), the debounced variant arms on a single negative running sum, then fires a tick and disarms after $k=\max(1,\operatorname{round}(\sqrt{n_w}))$ consecutive positive running sums.}
\label{tab:settings}
\begin{ruledtabular}
\begin{tabular}{lccc}
Setting & $c_w$ & $c$ & debounce\\\hline
1--3 & $1/8$ & $1,2,3$ & no\\
4 (main text) & $1/4$ & $1$ & no\\
5, 6 & $1/4$ & $2,3$ & no\\
7--9 & $1/2$ & $1,2,3$ & no\\
10 & $1/4$ & $0$ & yes\\
\end{tabular}
\end{ruledtabular}
\end{table}

\begin{table}[t]
\caption{Extrapolations of the precision to $N=1600$ (units of $\kappa$) from $N\le800$. The intervals propagate the uncertainty of the fitted data only.}
\label{tab:extrap}
\begin{ruledtabular}
\begin{tabular}{lcc}
Form & value & interval\\\hline
power law & $127.0$ & $[78.3,281.4]$\\
$S/\log S$ & $120.6$ & $[100.8,156.9]$\\
plateau & $51.7$ & $[33.4,111.4]$\\
\end{tabular}
\end{ruledtabular}
\end{table}

\begin{figure*}[!htbp]
\centering
\includegraphics[width=\textwidth]{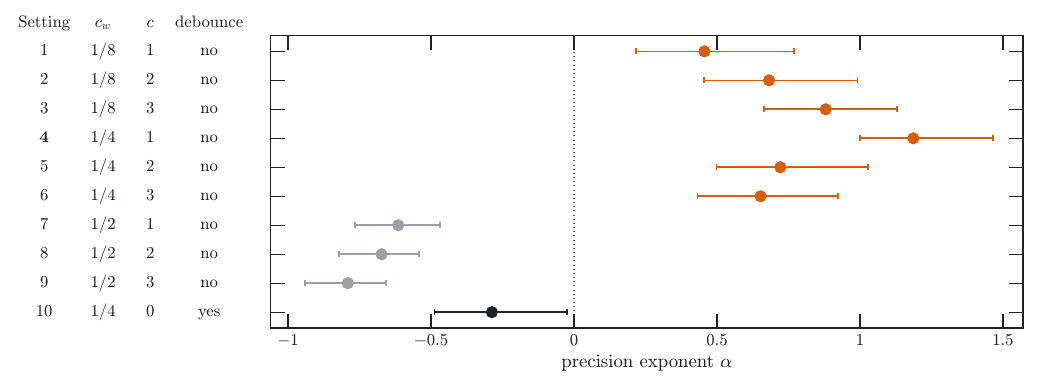}
\caption{Exponents of the precision for the ten decoder settings of Table~\ref{tab:settings} (zero temperature, $r=2$, $N\ge100$), with $95\%$ bootstrap intervals. Setting~4 is used in the main text; gray, the widest windows ($c_w=1/2$); black, the debounced variant.}
\label{fig:settings}
\end{figure*}

\textit{All settings.}---Figure~\ref{fig:settings} shows the exponent of every setting. All six settings with $c_w\le1/4$ and no debouncing have positive exponents. The widest windows ($c_w=1/2$) produce extra sign changes within a cycle, and the debounced variant adds extra ticks relative to the once-per-cycle reference; both give negative exponents.

\textit{Mean subtraction.}---A decoder that subtracts the mean label imbalance from the running sum and uses thresholds proportional to the window length fails. Cycles on small orbits, whose label contrast is weak, then cannot cross both thresholds, so that the number of registered cycles depends on the orbit. With its setting chosen in the same way, its precision exponent is $-1.90$ $[-2.13,-1.67]$ for $N=100$--$800$ at $r=2$.

\begin{figure*}[!htbp]
\centering
\includegraphics[width=\textwidth]{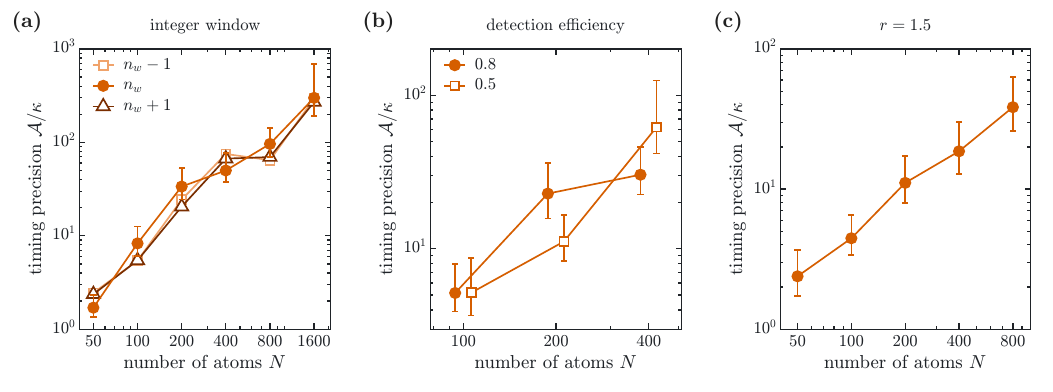}
\caption{Robustness of the phase clock on the same records: (a) integer window $n_w-1$, $n_w$, $n_w+1$; (b) detection efficiencies $0.8$ and $0.5$; (c) $r=1.5$. Error bars, $95\%$ bootstrap intervals, in (a) for the window $n_w$ only.}
\label{fig:robust}
\end{figure*}

\textit{Robustness.}---The same decoder was applied unchanged to records with the integer window shifted by $\pm1$, to records thinned to detection efficiencies $0.8$ and $0.5$ ($N=100$--$400$), and to records at $r=1.5$ (exponent $1.01$ $[0.76,1.27]$ for $100\le N\le800$) [Fig.~\ref{fig:robust}]. Shuffling the labels within each record keeps the click times and the port totals but destroys their order. The decoder then gives fewer ticks at every size and almost none at large $N$ (Table~\ref{tab:shuffle}).

\textit{Estimator check.}---For the total photon count the exact precision is known, which tests the trajectory estimates in Fig.~3(c,d) of the main text. The estimates lie above the exact values at $8$ of the $11$ sizes, by up to $44\%$, and outside their $95\%$ intervals at zero temperature for $N=200$ (exact $1.28\,\kappa$, interval $[1.31,2.88]\,\kappa$) and $N=1600$ ($0.162\,\kappa$, $[0.172,0.359]\,\kappa$). The late-window fit is not the cause: applied to the exact finite-time variance at $N=200$, transient included, it returns the exact value, $1.2767\,\kappa$, to a relative difference below $10^{-7}$. The precision is the inverse of an estimated variance slope, which sampling noise biases upward (Jensen's inequality). Deviations of a few tens of percent are therefore the uncertainty of all trajectory estimates, including those of the phase clock; they are small compared with the three orders of magnitude between the phase clock and photon counting.

\begin{table}[t]
\caption{Number of decoded ticks over the full records ($0\le t\le30\,\tau_{\rm slow}$), using the same decoder as the precision analysis, before and after shuffling the labels within each record. At $N=1600$ only the first 64 records are included.}
\label{tab:shuffle}
\begin{ruledtabular}
\begin{tabular}{rrrrr}
 & \multicolumn{2}{c}{zero temperature} & \multicolumn{2}{c}{thermal}\\
$N$ & original & shuffled & original & shuffled\\\hline
50 & 26\,568 & 10\,526 & 21\,265 & 11\,876\\
100 & 58\,588 & 18\,805 & 44\,894 & 14\,087\\
200 & 120\,169 & 18\,373 & 91\,172 & 8\,289\\
400 & 241\,086 & 5\,769 & 183\,633 & 3\,879\\
800 & 861\,188 & 2\,596 & 368\,223 & 564\\
1600 & 241\,952 & 9 & -- & --\\
\end{tabular}
\end{ruledtabular}
\end{table}

\section{Phase locking and the error budget}

\textit{Definitions.}---Along each record we propagate the conditional state and store it at every click and at 64 equally spaced times per period. The crossing times of $m_y=1/r$ obtained from the stored states agree with those found during the simulation to within $8\times10^{-10}\,\kappa^{-1}$. From its first moments we form $w$ of Eq.~(2) and the orbit radius $\rho=|w|$; $|q|=\sqrt{1-\rho^2}$ labels the mean-field orbit of the same radius. The model, both detectors, and the initial states are symmetric under the reflection $m_x\to-m_x$. A record therefore cannot distinguish $q$ from $-q$: the conditional states have $\langle S_x\rangle_c=0$, and only $|q|$ is resolved. $W(t)$ is the net number of windings of $\arg w$, counted with shortest-arc phase jumps. $C(t)$ is the number of decoded ticks, and $E=C-W$ is the decoder error. A cycle runs from one reference crossing to the next. Reference crossings are the crossings of $m_y=1/r$ from below, each counted only after a crossing from above, which suppresses repeated counts from jitter.

\textit{Locking.}---The mean resultant length of the spin phase at the ticks is $0.959$ $[0.956,0.962]$, $0.982$ $[0.981,0.983]$, and $0.991$ $[0.990,0.991]$ at $N=50$, $200$, and $800$, with mean offsets of $0.690$, $0.665$, and $0.657$ cycles after the crossing of $m_y=1/r$ from below. 
At $N=800$, orbits with $|q|\ge0.95$ ($\rho\le0.31$) hold $85.3\%$ of the missed cycles and $4.3\%$ of the time at zero temperature, and $84.1\%$ and $4.1\%$ in the thermal model.

\begin{figure}[!htbp]
\centering
\includegraphics[width=\columnwidth]{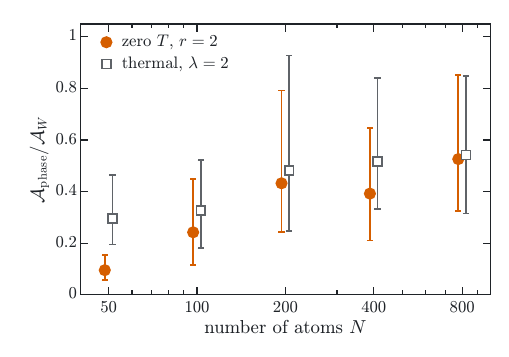}
\caption{Precision of the phase clock relative to the spin winding $W$ on the same records, with paired bootstrap intervals. In the thermal model $W$ uses the full unraveling, including the undetected absorptions.}
\label{fig:ratio}
\end{figure}

\textit{Error budget.}---With all covariances, ${\rm Var}\,C={\rm Var}\,W+{\rm Var}\,E+2\,{\rm Cov}(W,E)$, and the covariance is negative. To locate the noise, each cycle's centered increments $\delta X_k$ ($X=W,C,E$) are assigned to the radius bin $b$ of the cycle's start, so that $X=\sum_bX_b$ and ${\rm Var}\,X=\sum_b{\rm Cov}(X_b,X)$ exactly. The shares ${\rm Cov}(X_b,X)/{\rm Var}\,X$ are signed and need not be monotonic; they are not independent or causal fractions. Table~\ref{tab:budget} gives the share carried by the smallest orbits, which occupy $10\%$ of the time, together with the long-time correlation of $C$ and $W$. Figure~\ref{fig:ratio} shows the ratio of the precision of the phase clock to that of $W$. It rises from $0.10$ at $N=50$ to $0.53$ at $N=800$ at zero temperature, and from $0.30$ to $0.54$ in the thermal model, while the precision of $W$ grows roughly in proportion to $N$ (fitted exponents $0.87$ and $1.08$ for $N\ge100$), as expected if the spin phase diffuses at a rate of order $\kappa/N$. The exponents of the phase clock above one, $1.19$ and $1.30$, therefore describe the decoder approaching $W$ within the simulated range, not an asymptotic law.

\begin{table}[t]
\caption{Share (\%) of the long-time variance of the decoder error $E=C-W$ and of the winding $W$ carried by cycles that start on the smallest orbits (10\% of the time), and the long-time correlation of $C$ and $W$. Brackets, $95\%$ bootstrap intervals; at $N=1600$ only the first 64 records are used.}
\label{tab:budget}
\begin{ruledtabular}
\begin{tabular}{lccc}
$N$ & share of $E$ & share of $W$ & corr$(C,W)$\\\hline
\multicolumn{4}{l}{zero temperature, $r=2$}\\
50 & 29 [21, 37] & 57 [37, 75] & $-0.12$\\
100 & 38 [22, 54] & 51 [29, 64] & $0.34$\\
200 & 92 [80, 103] & 70 [52, 85] & $0.23$\\
400 & 99 [95, 104] & 66 [48, 80] & $0.09$\\
800 & 98 [93, 103] & 70 [59, 78] & $0.37$\\
1600 & 99 [77, 120] & 67 [34, 96] & $0.32$\\
\multicolumn{4}{l}{thermal, $\lambda=2$}\\
50 & 21 [8, 32] & 51 [39, 63] & $0.14$\\
100 & 47 [34, 58] & 59 [40, 73] & $0.22$\\
200 & 92 [82, 101] & 68 [54, 82] & $0.09$\\
400 & 100 [97, 104] & 79 [66, 90] & $0.46$\\
800 & 98 [93, 104] & 64 [48, 80] & $0.26$\\
\end{tabular}
\end{ruledtabular}
\end{table}

\section{Data availability}

The data underlying all figures, the results of the exact numerical calculations, and the simulation parameters and random seeds are openly available on Zenodo~\cite{HeDuan2026Data}. The full simulated photon records are available from the authors upon reasonable request.

\clearpage
\bibliography{refs}